# DiffDP: Radiotherapy Dose Prediction via a Diffusion Model

Zhenghao Feng[1], Lu Wen[1], Peng Wang[1], Binyu Yan[1], Xi Wu[2], Jiliu Zhou[1,2], Yan Wang[1(✉)]

[1] School of Computer Science, Sichuan University, Chengdu, China
[2] School of Computer Science, Chengdu University of Information Technology, Chengdu, China

**Abstract.** Currently, deep learning (DL) has achieved the automatic prediction of dose distribution in radiotherapy planning, enhancing its efficiency and quality. However, existing methods suffer from the over-smoothing problem for their commonly used $L_1$ or $L_2$ loss with posterior average calculations. To alleviate this limitation, we innovatively introduce a diffusion-based dose prediction (DiffDP) model for predicting the radiotherapy dose distribution of cancer patients. Specifically, the DiffDP model contains a forward process and a reverse process. In the forward process, DiffDP gradually transforms dose distribution maps into Gaussian noise by adding small noise and trains a noise predictor to predict the noise added in each timestep. In the reverse process, it removes the noise from the original Gaussian noise in multiple steps with the well-trained noise predictor and finally outputs the predicted dose distribution map. To ensure the accuracy of the prediction, we further design a structure encoder to extract anatomical information from patient anatomy images and enable the noise predictor to be aware of the dose constraints within several essential organs, i.e., the planning target volume and organs at risk. Extensive experiments on an in-house dataset with 130 rectum cancer patients demonstrate the superiority of our method.

**Keywords:** Radiotherapy Treatment, Dose Prediction, Diffusion Model, Deep Learning.

## 1 Introduction

Radiotherapy, one of the mainstream treatments for cancer patients, has gained notable advancements in past decades. For promising curative effect, a high-quality radiotherapy plan is demanded to distribute sufficient dose of radiation to the planning target volume (PTV) while minimizing the radiation hazard to organs at risk (OARs). To achieve this, radiotherapy plans need to be manually adjusted by the dosimetrists in a trial-and-error manner, which is extremely labor-intensive and time-consuming [1]. Additionally, the quality of treatment plans might be variable among radiologists due to their different expertise and experience [2]. Consequently, it is essential to develop a robust methodology to automatically predict the dose distribution for cancer patients, relieving the burden on dosimetrists and accelerating the radiotherapy procedure.

---

[1] H. Feng and L. Wen—The authors contribute equally to this work.

Recently, the blossom of deep learning (DL) has given rise to a series of DL-based methods for the dose prediction task [3-8]. For example, Nguyen *et al.* [3] modified the traditional 2D UNet [9] to fulfill dose prediction for prostate cancer patients. Liu *et al.* [5] constructed a cascade 3D (C3D) UNet to combine the global and local anatomical features and made a coarse-to-fine prediction of the dose distribution for H&N cancer patients. Wang *et al.* [6] utilized a progressive refinement UNet (PRUNet) to refine the predictions from low resolution to high resolution. Besides the above UNet-based frameworks, Song *et al.* [7] employed the deepLabV3+ [10] to excavate contextual information from different scales, thus obtaining accuracy improvements in the dose prediction of rectum cancer. Mahmood *et al.* [8] utilized a generative adversarial network (GAN)-based method to predict the dose maps of oropharyngeal cancer.

Although the above methods have achieved good performance in predicting dose distribution, they suffer from the over-smoothing problem. These DL-based dose prediction methods always apply the $L_1$ or $L_2$ loss to guide the model optimization which calculates a posterior mean of the joint distribution between the predictions and the ground truth, leading to the over-smoothed predicted images without important high-frequency details [11]. We display predicted dose maps from multiple deep models in Fig. 1. As shown, compared with the ground truth, i.e., (5) in Fig. 1, the predictions from (1) to (3) are blurred with fewer high-frequency details, such as ray shapes. These high-frequency features formed by ray penetration reveal the ray directions and dose attenuation with the aim of killing the cancer cells while protecting the OARs as much as possible, which are critical for radiotherapy. Consequently, exploring an automatic method to generate high-quality predictions with rich high-frequency information is important to improve the performance of dose prediction.

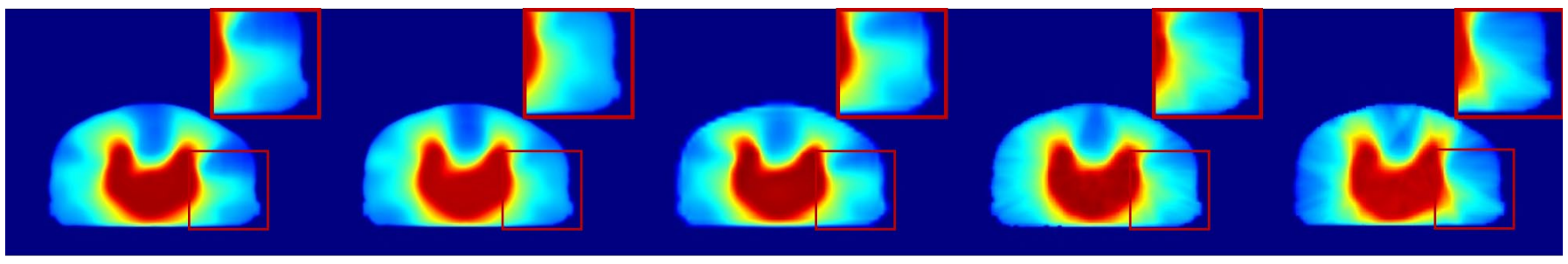

**Fig. 1.** Instances from a rectum cancer patient. (1) ~ (4): Dose maps predicted by UNet, GAN, deepLabV3+, and diffusion model.

Currently, diffusion model [12] has verified its remarkable potential in modeling complex image distributions in some vision tasks [13-14]. Unlike other DL models, the diffusion model is trained without any extra assumption about target data distribution, thus evading the average effect and alleviating the over-smoothing problem [15]. Fig.1 (4) provides an example in which the diffusion-based model predicts a dose map with shaper and clearer boundaries of ray-penetrated areas. Therefore, introducing a diffusion model to the dose prediction task is a worthwhile endeavor.

In this paper, we investigate the feasibility of applying a diffusion model to the dose prediction task and propose a diffusion-based model, called DiffDP, to automatically predict the clinically acceptable dose distribution for rectum cancer patients. Specifically, the DiffDP consists of a forward process and a reverse process. In the forward process, the model employs a Markov chain to gradually transform dose distribution maps with complex distribution into Gaussian distribution by progressively adding predefined noise. Then, in the reverse process, given a pure Gaussian noise, the model gradually removes the noise in multiple steps and finally outputs the predicted dose map. In this procedure, a noise predictor is trained to predict the noise added in the corresponding step of the forward process. To further ensure the accuracy of the predicted dose distribution for both the PTV and OARs, we design a DL-based structure encoder to extract the anatomical information from the CT image and the segmentation masks of the PTV and OARs. Such anatomical information can indicate the structure and relative position of organs. By incorporating the anatomical information, the noise predictor can be aware of the dose constraints among PTV and OARs, thus distributing more appropriate dose to them and generating more accurate dose distribution maps.

Overall, the contributions of this paper can be concluded as follows: (1) We propose a novel diffusion-based model for dose prediction in cancer radiotherapy to address the over-smoothing issue commonly encountered in existing DL-based dose prediction methods. *To the best of our knowledge, we are the first to introduce the diffusion model for this task.* (2) We introduce a structure encoder to extract the anatomical information available in the CT images and organ segmentation masks, and exploit the anatomical information to guide the noise predictor in the diffusion model towards generating more precise predictions. (3) The proposed DiffDP is extensively evaluated on a clinical dataset consisting of 130 rectum cancer patients, and the results demonstrate that our approach outperforms other state-of-the-art methods.

## 2 Methodology

An overview of the proposed diffDP model is illustrated in Fig. 2, containing two Markov chain processes: a forward process and a reverse process. An image set of cancer patient is defined as $\{x, y\}$, where $x \in R^{H\times W\times(2+o)}$ represents the structure images, "2" signifies the CT image and the segmentation mask of the PTV, and $o$ denotes the total number of segmentation mask of OARs. Meanwhile, $y \in R^{H\times W\times 1}$ is the corresponding dose distribution map for $x$. Concretely, the forward process produces a sequence of noisy images $\{y_0, y_1, \dots, y_T\}, y_0 = y$ by gradually adding a small amount of noise to $y$ in $T$ steps with the noise increased at each step and a noise predictor $f$ is constructed to predict the noise added to $y_{t-1}$ by treating $y_t$, anatomic information from $x$ and embedding of step $t$ as input. To obtain the anatomic information, a structure encoder $g$ is designed to extract the crucial feature representations from the structure images. Then, in the reverse process, the model progressively deduces the dose distribution map by iteratively denoising from $y_T$ using the well-trained noise predictor.

### 2.1 Diffusion Model

The framework of DiffDP is designed following the Denoising Diffusion Probabilistic Models (DDPM) [16] which contains a forward process and a reverse process. By utilizing both processes, the DiffDP model can progressively transform the Gaussian noise into complex data distribution.

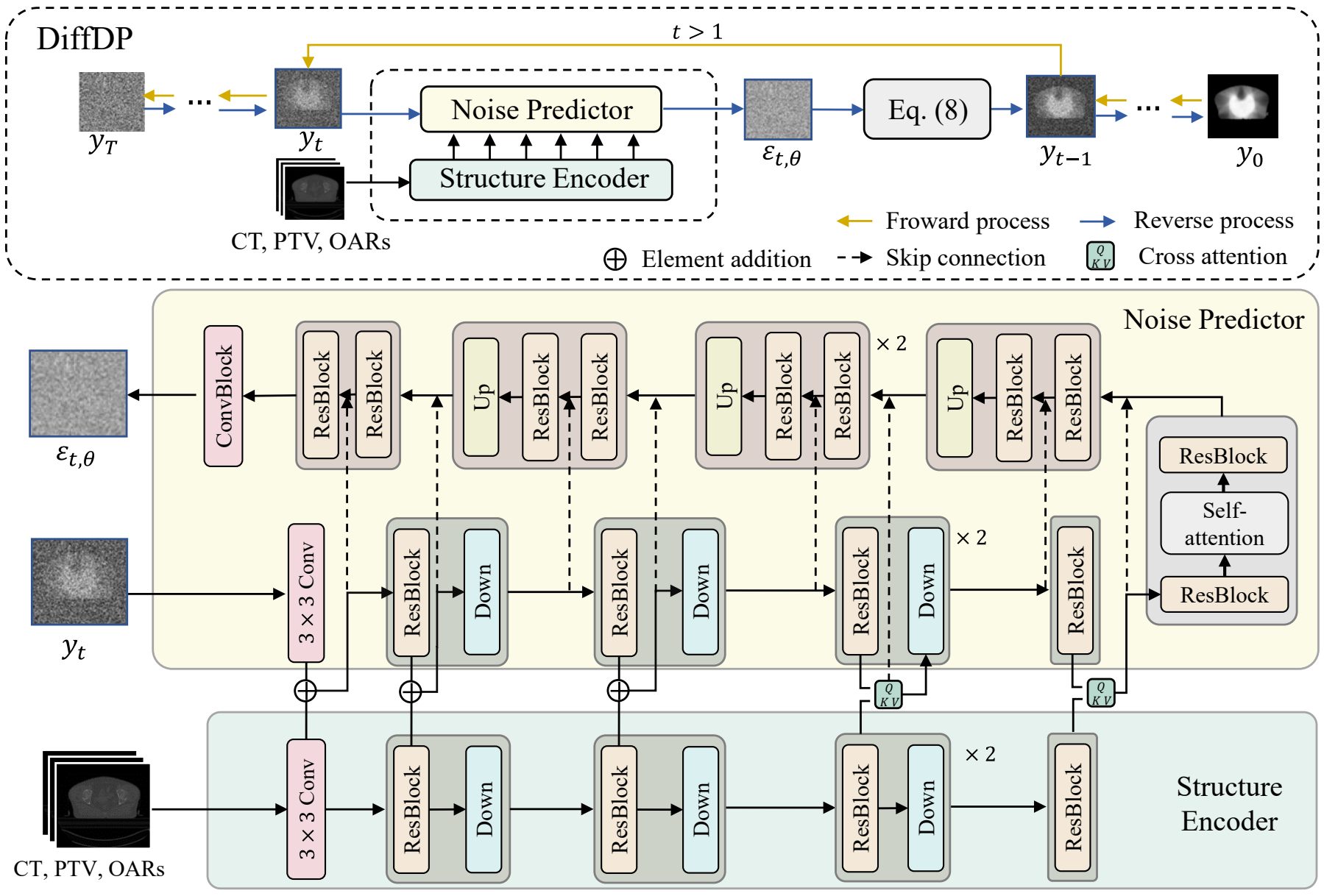

**Fig. 2.** Illustration of the proposed DiffDP network.

**Forward Process.** In the forward process, the DiffDP model employs the Markov chain to progressively add noise to the initial dose distribution map $y_0 \sim q(y_0)$ until the final disturbed image $y_T$ becomes completely Gaussian noise which is represented as $y_T \sim \mathcal{N}(y_T \mid 0, I)$. This forward process can be formulated as:

$$q(y_{1:T} \mid y_0) := \prod_{t=1}^{T} q(y_t \mid y_{t-1}), \tag{1}$$

$$q(y_t \mid y_{t-1}) := \mathcal{N}\left(y_t; \sqrt{\alpha_t} y_{t-1}, (1-\alpha_t)I\right), \tag{2}$$

where $\alpha_t$ is the unlearnable standard deviation of the noise added to $y_{t-1}$.

Herein, the $\alpha_t$ $(t = 1, \ldots, T)$ could accumulate during the forward process, which can be treated as the noise intensity $\gamma_t = \prod_{i=1}^{t} \alpha_i$. Based on this, we can directly obtain the distribution of $y_t$ at any step $t$ from $y_0$ through the following formula:

$$q(y_t \mid y_0) = \mathcal{N}\left(y_t; \sqrt{\gamma_t} y_0, (1-\gamma_t)I\right), \tag{3}$$

where the disturbed image $y_t$ is sampled using:

$$y_t = \sqrt{\gamma_t} y_0 + \sqrt{1-\gamma_t}\varepsilon_t, \tag{4}$$

in which $\varepsilon_t \sim \mathcal{N}(0, I)$ is random noise sampled from normal Gaussian distribution.

**Reverse Process.** The reverse process also harnesses the Markov chain to progressively convert the latent variable distribution $p_\theta(y_T)$ into distribution $p_\theta(y_0)$ parameterized by $\theta$. Corresponding to the forward process, the reverse one is a denoising transformation under the guidance of structure images $x$ that begins with a standard Gaussian distribution $y_T \sim \mathcal{N}(y_T \mid 0, I)$. This reverse inference process can be formulated as:

$$p_\theta(y_{0:T} \mid y_t, x) = p(y_T) \prod_{t=1}^{T} p_\theta(y_{t-1} \mid y_t, x), \tag{5}$$

$$p_\theta(y_{t-1} \mid y_t, x) = \mathcal{N}(y_{t-1}; \mu_\theta(x, y_t, \gamma_t), \sigma_t^2 I). \tag{6}$$

where $\mu_\theta(x, y_t, t)$ is a learned mean, and $\sigma_t$ is a unlearnable standard deviation. Following the idea of [16], we parameterize the mean of $\mu_\theta$ as:

$$\mu_\theta(x, y_t, \gamma_t) = \frac{1}{\sqrt{\alpha_t}}\left(y_t - \frac{1-\alpha_t}{\sqrt{1-\gamma_t}}\varepsilon_{t,\theta}\right), \quad (7)$$

where $\varepsilon_{t,\theta}$ is a function approximator intended to predict $\varepsilon_t$ from the input $x$, $y_t$ and $\gamma_t$. Consequently, the reverse inference at two adjacent steps can be expressed as:

$$y_{t-1} \leftarrow \frac{1}{\sqrt{\alpha_t}}\left(y_t - \frac{1-\alpha_t}{\sqrt{1-\gamma_t}}\varepsilon_{t,\theta}\right) + \sqrt{1-\alpha_t}z_t, \quad (8)$$

where $z_t \sim \mathcal{N}(0, I)$ is a random noise sampled from normal Gaussian distribution.

### 2.2 Structure Encoder

Vanilla diffusion model has difficulty preserving essential structural information and produce unstable results when predicting dose distribution maps directly from noise with a simple condition mechanism. To address this, we design a structure encoder $g$ that effectively extracts the anatomical information from the structure images guiding the noise predictor to generate more accurate dose maps by incorporating extracted structural knowledge. Concretely, the structure encoder includes five operation steps, each with a residual block (ResBlock) and a Down block, except for the last one. The ResBlock consists of two convolutional blocks (ConvBlock), each containing a 3×3 convolutional (Conv) layer, a GroupNorm (GN) layer, and a Swish activation function. The residual connections are reserved for preventing gradient vanishment in the training. The Down block includes a 3×3 Conv layer with a stride of 2. It takes structure image $x$ as input, which includes the CT image and segmentation masks of PTV and OARs, and evacuates the compact feature representation in different levels to improve the accuracy of dose prediction. The structure encoder is pre-trained by $L_1$ loss and the corresponding feature representation $x_e = g(x)$ is then fed into the noise predictor.

### 2.3 Noise Predictor

The purpose of the noise predictor $f(x_e, y_t, \gamma_t)$ is to predict the noise added on the distribution map $y_t$ with the guidance of the feature representation $x_e$ extracted from the structure images $x$ and current noise intensity $\gamma_t$ in each step $t$. Inspired by the great achievements of UNet [9], we employ a six-level UNet to construct the noise predictor. Specifically, the encoder holds the similar architecture with the structure encoder while the decoder comprises five deconvolution blocks to fulfill the up-sampling operation, and each contains an Up block and two ResBlock, except for the last one which discards the UP block. In each Up block, the Nearest neighbor up-sampling and a Conv layer with a kernel size of 1 are used. A bottleneck with two Resblocks and a self-attention module is embedded between the encoder and decoder.

In the encoding procedure, to guide the noise predictor with essential anatomical structure, the feature representations respectively extracted from the structure images $x$ and noisy image $y_t$ are simultaneously fed into the noise predictor. Firstly, $y_t$ is encoded into feature maps through a convolutional layer. Then, these two feature maps are fused by element-wise addition, allowing the structure information in $x$ to be transferred to the noise predictor. The following two down-sampling operations retain the addition operation to complete information fusion, while the last three use a cross-attention mechanism to gain similarity-based structure guidance at deeper levels.

In the decoding procedure, the noise predictor restores the feature representations captured by the encoder to the final output, i.e., the noise $\varepsilon_{t,\theta} = f(x_e, y_t, \gamma_t)$ in step $t$. The skip connections between the encoder and decoder are reserved for multi-level feature reuse and aggregation.

### 2.4 Objective Function

The main purpose of the DiffDP model is to train the noise predictor $f$ and structure encoder $g$, so that the predicted noise $\varepsilon_{t,\theta} = f(g(x), y_t, \gamma_t)$ in the reverse process can approximate the added noise $\varepsilon_t$ in the forward process. To achieve this, we define the objective function as:

$$\min_{\theta} \mathbb{E}_{(x,y)} \mathbb{E}_{\varepsilon,\gamma} \left\| f\left( g(x), \underbrace{\sqrt{\gamma} y_0 + \sqrt{1-\gamma} \varepsilon_t}_{y_t}, \gamma_t \right) - \varepsilon_t \right\|, \varepsilon_t \sim \mathcal{N}(0, I) \quad (9)$$

For a clearer understanding, the training procedure is summarized in Algorithm 1.

**Algorithm 1:** Training procedure

1: **Input:** Input image pairs $P = \{(x_i, y_i)\}_{i=1}^{I}$ where $x$ is the structure image and $y$ is the corresponding dose distribution map, the total number of diffusion steps $T$.
2: **Initialize:** Randomly initialize the noise predictor $f$ and pre-trained structure encoder $g$.
3: **Repeat**
4: Sample $(x, y) \sim P$
5: Sample $\varepsilon_t \sim \mathcal{N}(0, I)$, and $t \sim Uniform(\{1, \dots, T\})$
6: Perform the gradient step on Equation (9)
7: **until** converged

### 2.5 Training Details

We accomplish the proposed network in the PyTorch framework. All of our experiments are conducted through one NVIDIA RTX 3090 GPU with 24GB memory and a batch size of 16 with an Adaptive moment estimation (Adam) optimizer. We train the whole model for 1500 epochs (about 1.5M training steps) where the learning rate is initialized to 1e-4 and reset to 5e-5 after 1200 epochs. The parameter $T$ is set to 1000. Additionally, the noise intensity is initialized to 1e-2 and decayed to 1e-4 linearly along with the increase of steps.

## 3. Experiments and Results

**Dataset and Evaluations.** We measure the performance of our model on an in-house rectum cancer dataset which contains 130 patients who underwent volumetric modulated arc therapy (VMAT) treatment at West China Hospital. Concretely, for every patient, the CT images, PTV segmentation, OARs segmentations, and the clinically planned dose distribution are included. Additionally, there are four OARs of rectum cancer containing the bladder, femoral head R, femoral head L, and small intestine. We

**Table 1.** Quantitative comparison results with state-of-the-art methods in terms of $\Delta HI$, $\Delta D_{98}$, $\Delta D_2$, and $\Delta D_{max}$. * means our method is significantly better than compared method with p < 0.05 via paired t-test.

| Methods | $\Delta HI$ | $\Delta D_{98}$ | $\Delta D_2$ | $\Delta D_{max}$ |
|---|---|---|---|---|
| UNet [3] | 0.0494(5.8E-3)* | 0.0428(5.2E-3)* | 0.0048(1.1E-5)* | 0.0186(2.6E-5)* |
| GAN [8] | 0.0545(5.2E-3)* | 0.0431(4.3E-3)* | 0.0253(2.0E-5)* | 0.0435(3.5E-6)* |
| deepLabV3+ [7] | 0.0448(4.8E-3)* | 0.0416(4.2E-3) | 0.0036(7.8E-6)* | 0.0139(8.2E-6)* |
| C3D [5] | 0.0460(5.6E-3)* | 0.0400(4.9E-3) | 0.0077(1.8E-5)* | 0.0206(3.0E-5)* |
| PRUNet [6] | 0.0452(5.2E-3)* | 0.0407(4.5E-3) | 0.0088(6.2E-5)* | 0.0221(4.3E-5)* |
| Proposed | **0.0413(4.5E-3)** | **0.0392(4.1E-3)** | **0.0008(1.1E-5)** | **0.0005(4.4E-6)** |

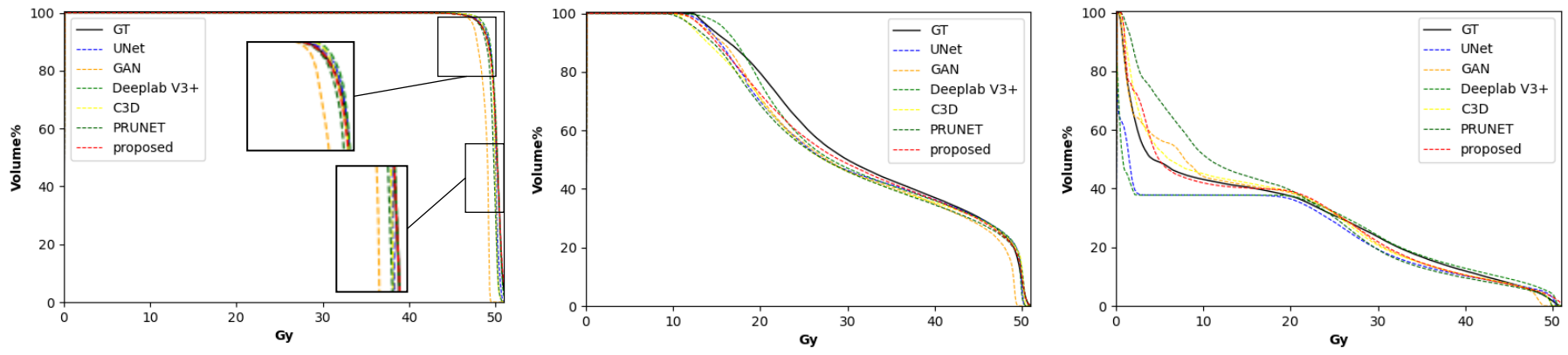

PTV Bladder Small intestine

**Fig. 3.** Visual comparison of DVH curves by our method and SOTA methods, including DVH curves of PTV, bladder, and small intestine.

randomly select 98 patients for model training, 10 patients for validation, and the remaining 22 patients for test. The thickness of the CTs is 3mm and all the images are resized to the resolution of 256×256 before the training procedure.

We measure the performance of our proposed model with multiple metrics. Considering Dm represents the minimal absorbed dose covering m% percentage volume of PTV, we involve $D_{98}$, $D_2$, maximum dose ($D_{max}$), and mean dose ($D_{mean}$) as metrics. Besides, the heterogeneity index (HI) is used to quantify dose heterogeneity [17]. To quantify performance more directly, we calculate the difference (Δ) of these metrics between the ground truth and the predicted results. More intuitively, we involve the dose volume histogram (DVH) [18] as another essential metric of dose prediction performance. When the DVH curves of the predictions are closer to the ground truth, we can infer higher prediction accuracy.

**Comparison with State-Of-The-Art Methods.** To verify the superior accuracy of our proposed model, we select multiple state-of-the-art (SOTA) models in dose prediction, containing UNet (2017) [3], GAN (2018) [8], deepLabV3+ (2020) [7], C3D (2021) [5], and PRUNet (2022) [6], for comparison. The quantitative comparison results are listed in Table. 1 where our method outperforms the existing SOTAs in terms of all metrics. Specifically, compared with deepLabV3+ with the second-best accuracy in $\Delta HI$ (0.0448) and $\Delta D_{98}$ (0.0416), the results generated by the proposed are 0.0035 and 0.0014 lower, respectively. As for $\Delta D_2$ and $\Delta D_{max}$, our method gains overwhelming performance with 0.0008 and 0.0005, respectively. Moreover, the paired t-test is conducted to investigate the significance of the results. The p-values between the proposed and other SOTAs are almost all less than 0.05, indicating that the enhancement of performance is statistically meaningful.

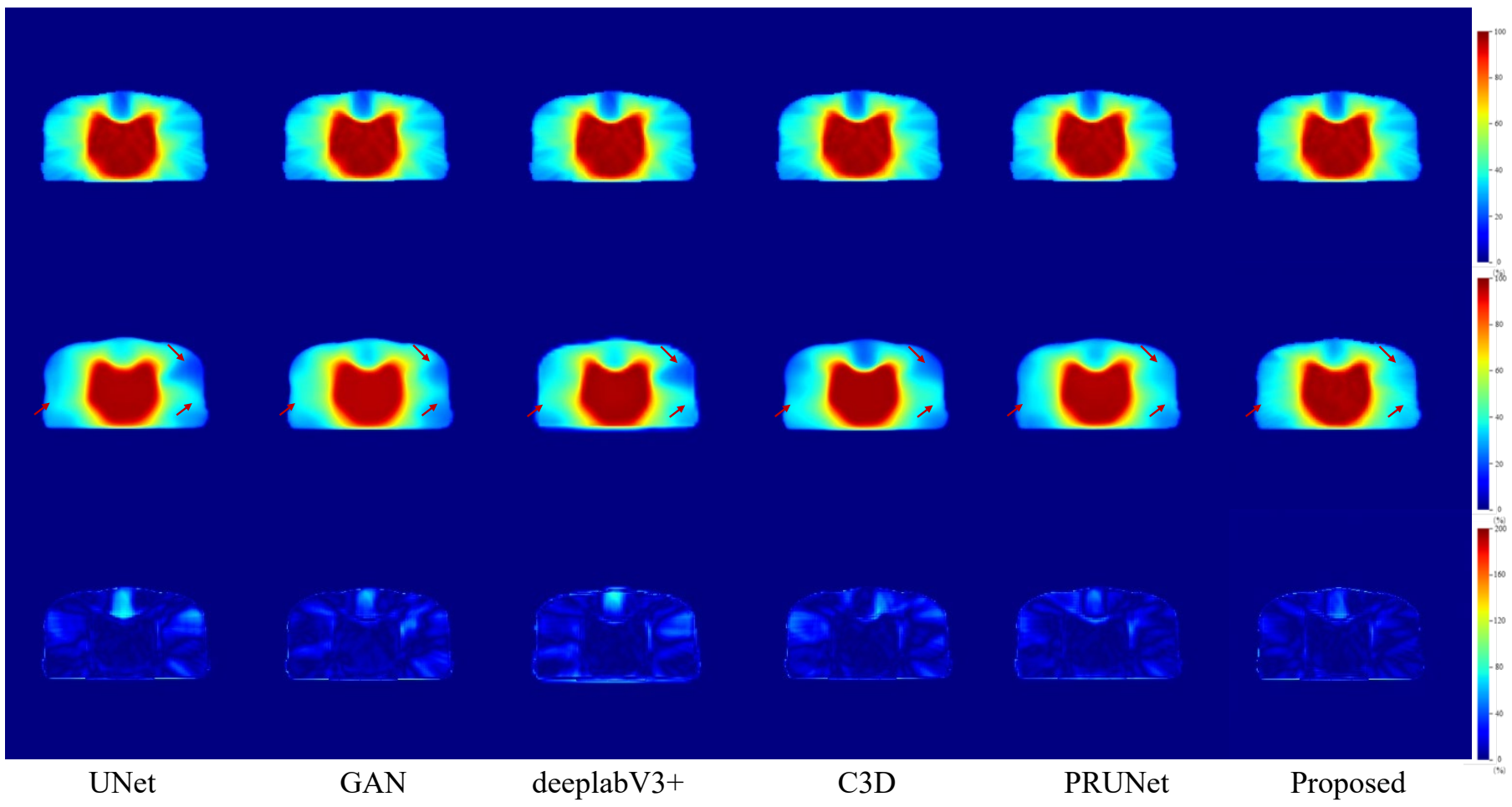

**Fig. 4.** Visual comparison with SOTA models. From top to bottom: ground truth, predicted dose map and corresponding error maps.

Besides the quantitative results, we also present the DVH curves derived by compared methods in Fig. 3. The results are compared on PTV as well as two OARs: bladder and small intestine. Compared with other methods, the disparity between the DVH curves of our method and the ground truth is the smallest, demonstrating the superior performance of the proposed.

Furthermore, we display the visualization comparison in Fig.4. As we can see, the proposed model achieves the best visual quality with clearer and sharper high-frequency details (as indicated by red arrows). Furthermore, the error map of the proposed is the darkest, suggesting the least disparity compared with the ground truth.

**Ablation Study.** To study the contributions of key components of the proposed method, we conduct the ablation experiments by 1) removing the structure encoder from the proposed method and concatenating the anatomical images $x$ and noisy image $y_t$ together as the original input for diffusion model (denoted as Baseline); 2) the proposed DiffDP model. The quantitative results are given in Table 2. We can clearly see the performance for all metrics is enhanced with the structure encoder, demonstrating its effectiveness in the proposed model.

**Table 2.** Ablation study of our method in terms of $\Delta HI$, $\Delta D_{98}$, $\Delta D_2$, and $\Delta D_{mean}$. * means our method is significantly better than other variants with $p < 0.05$ via paired t-test.

| Methods | $\Delta HI$ | $\Delta D_{98}$ | $\Delta D_2$ | $\Delta D_{mean}$ |
|---|---|---|---|---|
| Baseline | 0.0444(4.7E-3)* | 0.0426(4.2E-3) | 0.0021(1.1E-5)* | 0.0246(7.5E-4)* |
| Proposed | **0.0413(4.5E-3)** | **0.0392(4.1E-3)** | **0.0008(1.1E-5)** | **0.0154(6.5E-4)** |

## 4. Conclusion

In this paper, we introduce a novel diffusion-based dose prediction (DiffDP) model for predicting the radiotherapy dose distribution of cancer patients. The proposed method involves a forward and a reverse process to generate accurate prediction by progressively transferring the Gaussian noise into a dose distribution map. Moreover, we propose a structure encoder to extract anatomical information from patient anatomy images and enable the model to concentrate on the dose constraints within several essential organs. Extensive experiments on an in-house dataset with 130 rectum cancer patients demonstrate the superiority of our method.

## Acknowledgements

This work is supported by the National Natural Science Foundation of China (NSFC 62071314), Sichuan Science and Technology Program 2023YFG0263, 2023NSFSC0497, 22YYJCYJ0086, and Opening Foundation of Agile and Intelligent Computing Key Laboratory of Sichuan Province.